# Creation of moiré bands in a monolayer semiconductor by spatially periodic dielectric screening

Yang Xu[1], Connor Horn[1], Jiacheng Zhu[1], Yanhao Tang[1], Liguo Ma[1], Lizhong Li[1], Song Liu[2], Kenji Watanabe[3], Takashi Taniguchi[3], James C. Hone[2], Jie Shan[1,4,5*], Kin Fai Mak[1,4,5*]

[1]School of Applied and Engineering Physics, Cornell University, Ithaca, NY, USA
[2]Department of Mechanical Engineering, Columbia University, New York, NY, USA
[3]National Institute for Materials Science, 1-1 Namiki, 305-0044 Tsukuba, Japan
[4]Laboratory of Atomic and Solid State Physics, Cornell University, Ithaca, NY, USA
[5]Kavli Institute at Cornell for Nanoscale Science, Ithaca, NY, USA
Email: jie.shan@cornell.edu; kinfai.mak@cornell.edu

**Moiré superlattices of two-dimensional van der Waals materials have emerged as a powerful platform for designing electronic band structures and discovering emergent physical phenomena [1–16]. A key concept involves the creation of long-wavelength periodic potential and moiré bands in a crystal through interlayer hybridization when two materials are overlaid. Here we demonstrate a new approach based on spatially periodic dielectric screening to create moiré bands in a monolayer semiconductor. It relies on reduced dielectric screening of the Coulomb interactions in monolayer semiconductors and their environmental dielectric-dependent electronic band structure [17–21]. We observe optical transitions between moiré bands in monolayer $WSe_2$ when it is placed close to small angle-misaligned graphene on hexagonal boron nitride. The moiré bands are a result of long-range Coulomb interactions, strongly gate-tunable, and can have versatile superlattice symmetries independent of the crystal lattice of the host material. Our result also demonstrates that monolayer semiconductors are sensitive local dielectric sensors.**

A moiré pattern is formed when two materials of slightly different orientations or lattice constants are overlaid. The moiré pattern introduces a new length scale that is many times the lattice constant of the original materials for Bragg scattering of Bloch electrons in each layer. This gives rise to moiré bands and rich emergent phenomena, including the Hofstadter's butterfly physics [2–4], correlated insulating states [6–9], superconductivity [5], quantum magnetism [10–12], and moiré excitons [13–16]. Moiré bands in these structures are created by spatially periodic potential from interlayer electronic hopping (hybridization), whose strength depends on atomic registries between two materials. Such an interaction relies on wave function overlap of the atomic orbitals and is short-ranged, that is, suppressed exponentially with layer separation.

Monolayer semiconductors, such as transition metal dichalcogenides (TMDs) $MX_2$ (M = Mo, W; X = S, Se), present a new approach to create moiré bands. In these atomically thin materials dielectric screening of the Coulomb interactions is ineffective [17–21]. The electric-field lines between charges extend substantially outside the material. This leads to an electronic band structure that is strongly dependent on the surroundings [19,20]. A reduction of 100's meV in the quasiparticle band gap energy from that of a freestanding monolayer is possible by engineering the environmental dielectric function $\varepsilon$ [19–22]. If a spatially periodic $\varepsilon$ is introduced, the conduction and valence band edges shift with the same period in opposite directions (Fig. 1a). This provides long-wavelength potential for electrons and holes to form moiré bands. The method relies on spatially periodic dielectric screening of the Coulomb interactions, which are long-ranged and potentially also gate-tunable through $\varepsilon$. In addition, versatile superlattice symmetries that are independent of the crystal lattice of the host material can be designed through the dielectric substrate.

In this work, we demonstrate the creation of moiré bands in $WSe_2$ monolayers by environmental dielectric engineering. We use small angle-misaligned graphene on hexagonal boron nitride (hBN) as a two-dimensional (2D) periodic dielectric substrate. Because of a small difference in lattice constant, graphene and hBN with a small rotational misalignment form a hexagonal moiré superlattice (Fig. 1b). The superlattice period $\lambda$ is tunable by the misalignment angle [1]. Figure 1c shows the device schematic. A $WSe_2$ monolayer is placed close to a graphene layer angle-aligned to an hBN substrate. A thin hBN spacer (not shown, angle-misaligned with graphene) can be inserted to exclude any potential electronic hybridization between $WSe_2$ and graphene. The carrier density in the device can be varied by a bottom gate, which is made of an hBN dielectric and graphite electrode. Since the Dirac point of graphene is deep inside the $WSe_2$ band gap (Fig. 1d) [23], charges are injected only into the graphene layer for the entire range of experimentally accessible gate voltages. We examine the electronic band structure of $WSe_2$ by optical reflection spectroscopy. Details on device fabrication and optical measurements are described in Methods.

Figure 2a and 2d show the reflection contrast spectrum of $WSe_2$ as a function of gate voltage $V_g$ for device D1 and D2, both without a spacer. The graphene layer and the hBN substrate have a large rotational misalignment in device D1, and a small misalignment of about 0.6° in device D2. The graphene layer is charge neutral near $V_g = 0$ V. The carrier density (right axis of Fig. 2a, 2d) is evaluated from $V_g$ and the gate capacitance. The optical spectrum is dominated by the 1s exciton in $WSe_2$ near 1.71 eV. The 1s exciton exhibits a small redshift with doping in graphene. The negligible change in the oscillator strength and the absence of charged 1s exciton features verify that the $WSe_2$ layer remains charge neutral for all gate voltages [24,25].

In contrast, significant gate dependence is observed for the feature around 1.8 eV. It redshifts by nearly 40 meV and loses its oscillator strength with doping. In particular, it is approximately symmetric for electron and hole doping in device D1. Two representative spectra are shown in Fig. 2b. Near the graphene Dirac point ($V_g = 0$ V), the exciton effect is substantial with discernable 2s and 3s states that evolve into the band edge transitions with increasing photon energy. For heavily doped graphene ($V_g = 5$ V), the exciton excited states merge into the band edge transitions (or Fermi edge singularity) and are no longer identifiable[26].

In device D2 the feature around 1.8 eV is no longer symmetric for electron and hole doping (Fig. 2d). In addition to the strong spectral feature around the graphene Dirac point, a satellite feature emerges on the hole doping side. More intriguingly, two replicas appear 20-30 meV above the fundamental band edge transitions. All three transitions share similar doping dependences. Figure 2e illustrates two horizontal linecuts also at $V_g = 0$ V and 5 V. No replicas are observed for the 1s exciton within the experimental uncertainty of 0.1% in the reflection contrast. Two additional devices, D3 and D4, with different rotational misalignments between graphene and hBN are studied in Fig. 3a and 3b. They exhibit similar behaviors to those of device D2. However, the satellite feature occurs at different hole densities in different devices. There is also a positive correlation between the hole density, at which the satellite feature is observed, and the energy separation between the replicas and the fundamental band edge transitions.

To understand the origin of these new spectral features, we first determine the energies of the band-to-band transition (or Fermi edge singularity) in monolayer WSe$_2$. Near the Dirac point, we evaluate the band-to-band transition energy from the 2s and 3s exciton spacing using the 2D hydrogen model as demonstrated by previous studies [18,27,28]. The extracted values agree well with the analysis of a polarized magneto-optical study, in which we determine the band-to-band transition energies by fitting the magnetic-field-dependent diamagnetic shift of the exciton states using the non-hydrogenic Keldysh potential model [29]. Away from the Dirac point, the exciton excited states (up to 11s) can be identified under an out-of-plane magnetic field. They evolve into interband Landau level transitions under high magnetic fields. They are equally spaced, depend linearly on field, and merge into a single value at zero field, from which we determine the band edge transition energy. Details are provided in Methods.

Figure 2c and 2f summarize the extracted quasiparticle band gap of monolayer WSe$_2$ as a function of gate voltage in device D1 and D2, respectively. It decreases monotonically with increasing gate voltage in device D1. Away from the Dirac point, the quasiparticle gap scales approximately linearly with $|n|^{1/2}$ (inset, Fig. 2c). In contrast, the quasiparticle gap in device D2 shows a satellite peak on the hole doping side. Figure 2f also shows the gate-dependent transition energies of the replica features. Away from the Dirac points, the spacing between the replicas and the fundamental band edge transitions, $\Delta_1$ and $\Delta_2$ ($> \Delta_1$), are nearly doping independent.

The observed gate-dependent quasiparticle gap (and exciton binding energies) is a consequence of dynamical screening of the Coulomb interactions between charges in WSe$_2$ by graphene. In large angle-misaligned graphene on hBN, the dielectric function of graphene increases with doping monotonically and reduces both the quasiparticle gap and the exciton binding energies. Theoretical calculations have shown that the two effects are comparable and nearly cancel each other for the 1s exciton [20,21,30]. The 1s exciton is thus nearly doping independent. The exciton excited states merge into the red shifting band edge transitions rapidly. More quantitatively, the free carriers in graphene dominate its dielectric function through the 2D polarizability, which depends linearly on the electronic density of states in the long-wavelength limit, i.e. $|n|^{1/2}$ [31]. This agrees well with the observed doping dependence of the quasiparticle gap in Fig. 2c (see Methods for more details).

In small angle-misaligned graphene/hBN devices (Fig. 2f), the emergence of satellite features is a manifestation of the formation of moiré superlattices in graphene. New Dirac points are created where the k and – k bands are connected by Bragg scattering from the superlattice potential (Fig. 1b). A small energy gap is further opened at the primary Dirac point and the secondary Dirac point in the valence band by locally broken sublattice symmetry from structural relaxation of graphene [1,2,32]. The smaller dielectric function of graphene at these points weakens screening of the Coulomb interactions in $WSe_2$ and blueshifts the exciton excited states and the band edge transitions (Fig. 2d). Similar to the case of angle-misaligned graphene/hBN devices, the 1s exciton remains largely unaffected. The doping-dependent quasiparticle gap closely follows the electronic density of states of the graphene-hBN moiré superlattice [1,2], as expected for dielectric screening. We can also relate the density required to fill the secondary Dirac point, $n_{SDP} = \frac{8}{\sqrt{3}\lambda^2}$, to the superlattice period $\lambda$ (corresponding to four particles per unit cell including the spin and valley degeneracies).

With the dielectric-dependent quasiparticle gap in monolayer $WSe_2$ established, we discuss the origin of the optical transition replicas. The graphene/hBN moiré superlattices create spatially periodic electronic density of states and dielectric function [1]. This allows spatially periodic dynamical screening of Coulomb interactions and creates moiré bands in $WSe_2$ monolayers (Fig. 1a). New optical transitions become allowed between states that differ in momentum by reciprocal superlattice vectors. The optical transition replicas are compatible to the formation of moiré bands in monolayer $WSe_2$. In this picture, the first replica corresponds to transitions from the first hole moiré band to the second electron moiré band (Fig. 3f) or from the second hole moiré band to the first electron moiré band. It is expected at $\Delta_1 \approx \frac{2h^2}{3m\lambda^2}$ above the fundamental band edge transitions, where $h$ denotes Planck's constant and $m$ is the band mass of monolayer $WSe_2$. This explains the observed correlation between $n_{SDP}$ and $\Delta$ among different devices since they both scale as $1/\lambda^2$. The $1/\lambda^2$-dependence is a characteristic of a massive electron in $WSe_2$. The observation is not compatible with the picture of remote excitonic coupling to both single-particle excitations in graphene, in which a $1/\lambda$-dependence is expected due to the linear band dispersion in graphene, and collective plasmon excitations, in which a $1/\sqrt{\lambda}$-dependence is expected[31].

To further support this interpretation, we study the $\lambda$-dependence of $\Delta_1$ and $\Delta_2$ (away from the Dirac points) by fabricating different devices (Fig. 3e). We evaluate $\lambda$ of each device from the measured density difference between the primary and secondary Dirac points $n_{SDP}$. Both $\Delta_1$ and $\Delta_2$ depend on $1/\lambda^2$ linearly. From the slope of $\Delta_1$ we extract $m \sim 0.60 m_0$ ($m_0$ denoting the free electron mass). This value is close to the reported electron and hole masses (which are nearly identical) in monolayer $WSe_2$ [33]. The slope of $\Delta_2$ is about 1.3 times larger. We consider it unlikely the electron-hole asymmetry. Instead the value is close to the squared ratio of the first two reciprocal superlattice vectors along the $\Gamma K$ and $\Gamma M$ directions, $(2/\sqrt{3})^2$ (Fig. 3f). The second replica is therefore likely the same transition as the first replica along the $\Gamma K$ instead of the $\Gamma M$ direction. The intercepts in Fig. 3e (~ 2-3 meV) could provide an estimate for level anticrossings of the moiré bands. The value is compatible with the observed relative oscillator strength of the optical transition replicas compared to the fundamental band edge transitions. Further experimental studies are required to verify these assignments.

Finally, we demonstrate the long-range nature of the observed effect. We insert an hBN spacer of different thicknesses between the $WSe_2$ and graphene layer (Fig. 1c). Short-range electronic hybridization will be suppressed exponentially with spacer thickness. But the long-range dielectric screening effect is expected to decay gradually with spacer thickness, provided that it is smaller than the moiré period. Figure 3c is the gate-dependent reflection contrast spectrum of a part of device D3 with a one-layer hBN spacer (~ 0.36 nm). (The part with no spacer is shown in Fig. 3a). Figure 3d is the result for device D5 with a 6-layer hBN spacer (~ 2 nm). The characteristic secondary Dirac point and the optical transition replicas are observed in both devices. The energy separation between different moiré band transitions falls nicely onto the linear dependences in Fig. 3e (filled symbols). This indicates that the spacer does not alter the underlying physics. Meanwhile, with increasing spacer thickness (from Fig. 3a to 3c to 3d), the screening effect on the quasiparticle gap and the oscillator strength of the replicas decrease gradually. In particular, the second replica becomes not observable with 6-layer hBN spacer, reflecting a weaker moiré potential. All of these results support the picture of moiré band formation through the long-range dielectric screening.

In summary, we have demonstrated gate-tunable, long-range and spatially periodic dielectric screening of the Coulomb interactions in monolayer semiconductors. Moiré bands are formed as a result of the spatially periodic electronic band structure, and probed by optical spectroscopy. The optical transition replicas observed here bear certain similarity to the recently reported 1s moiré excitons in TMD heterostructures [13–15] that originate from the moiré exciton bands [34,35]. We do not observe 1s moiré excitons here. The 1s exciton sees almost no superlattice potential from the spatially periodic dielectric substrate because of the near perfect cancellation of the screening-renormalized quasiparticle gap and exciton binding energy [21,30]. This is consistent with the nearly environmental dielectric-independent 1s exciton energy. We have used graphene/hBN superlattices as a dielectric substrate. The approach can be generalized to other dielectric superlattices, including patterned ones with arbitrary superlattice symmetries [36], to tailor quasiparticle band structures and explore emergent phenomena in 2D semiconductors. Our work has also indicated that semiconducting TMD monolayers can serve as sensitive local dielectric sensors for 2D electronic systems.

## Methods

**Device fabrication and gating**
Atomically thin flakes of WSe$_2$, graphene, and hexagonal boron nitride (hBN) are first mechanically exfoliated from bulk crystals onto Si substrates and identified by their reflection contrast using an optical microscope. Van der Waals heterostructures are then prepared by a layer-by-layer dry-transfer method [37] and released onto Si substrates with a 285-nm oxide layer and pre-patterned gold electrodes.

Figure 1c illustrates the device schematic. Figure S1a shows an optical image of device D1. A typical device consists of WSe$_2$ and graphene monolayers (either directly touching or separated by a thin hBN spacer) that are sandwiched between two hBN flakes. The graphene monolayer extends out of the bottom hBN layer to contact a gold electrode. The hBN thickness is typically 20 to 40 nm. The bottom hBN and a few-layer graphite layer form a back gate. A gate voltage $V_g$ is applied to the few-layer graphite through a Keithley 2400 source meter while the graphene layer is grounded. The gate voltage varies the carrier density in graphene and screening of the Coulomb interactions of charges in WSe$_2$.

In order to produce small angle-misaligned graphene on hBN, we select flakes of graphene and hBN with sharp edges that are close to 30° or multiples of 30° and align the edges accordingly. An example is shown in Fig. S2d and S2f from device D3. These edges have a high probability of being the zigzag or armchair cleavage planes. We have fabricated 13 devices in total, among which 6 show moiré band transitions in monolayer WSe$_2$. The superlattice period $\lambda$ of these devices ranges from 9.4 to 14.7 nm.

**Determination of the hBN thickness and the superlattice period**
To verify the long-range nature of the spatially periodic dielectric screening effect, we have fabricated four devices with a thin hBN spacer ranging from one to six layers to separate the WSe$_2$ and graphene layers. We first identify thin hBN flakes on Si substrates by their optical reflection (see optical image of an hBN monolayer in Fig. S2c). We measure their thickness with an atomic force microscope (AFM, Park system XE7) on finalized devices. The measurements are performed first with the contact mode and verified with the tapping mode. An example is shown in Fig. S3. The contact mode also helps to clean the polymer residuals on the device surfaces.

The thickness of the bottom-gate dielectric hBN $t$ is also measured by AFM to determine the gate capacitance and the superlattice period $\lambda$. The gate capacitance per unit area is related to $t$ through $C_g = \varepsilon_r \varepsilon_0 / t$, where $\varepsilon_r \approx 3$ denotes the hBN out-of-plane dielectric constant [38] and $\varepsilon_0$ is the vacuum permittivity. For devices showing moiré band transitions, the voltage difference between the graphene primary Dirac point and the secondary Dirac point (SDP) is marked as $V_s$. It is related to the full moiré band filling density $n_{SDP} = C_g V_s / e$, where $e$ denotes the elementary charge. The period of the graphene/hBN superlattice is evaluated using $\frac{\sqrt{3}}{2} \lambda^2 = \frac{4}{n_{SDP}}$ for a hexagonal lattice, where the factor of 4 arises from the spin and valley degeneracies.

**Optical reflection contrast spectroscopy**
Optical measurements are performed with the devices loaded inside a close-cycle cryostat (Attocube, Attodry 1000) at base temperature of 3.5 K and under a magnetic field up to 9 T. A halogen lamp is used as a white light source, whose output is collected by a single-mode fiber and collimated by a 10× objective. The beam is focused onto the sample by an objective with a numerical aperture (N.A.) of 0.8. The beam diameter is about 1 μm and the power is below 1 nW on the devices. The reflected light is collected by the same objective and detected by a spectrometer equipped with a liquid-nitrogen cooled charge-coupled device (CCD) camera. The reflectance contrast ($\Delta R/R_0$) spectrum is obtained by comparing the reflected light spectrum from the sample ($R$) to that from the substrate right next to the sample ($R_0$) as $\Delta R/R_0 = (R-R_0)/R_0$. The substrate is featureless in the spectral region of interest. The measurement sensitivity for the reflection contrast is about 0.1%.

**Magneto-optical measurements and analysis**
Magneto-optical measurements are performed to help understand the gate-dependent optical transitions at zero magnetic field. In Fig. S4a and S5a, we show the magnetic-field dependence of the polarized reflection contrast spectrum of device D1 at $V_g$ = 0.045 V and 5 V, respectively. The out-of-plane magnetic field $B$ varies from -9 T to 9 T. The white light probe is left circularly polarized ($\sigma^+$). It is generated by a combination of a linear polarizer and an achromatic quarter-wave plate. The result for the right circularly polarized light ($\sigma^-$) is identical to that for $\sigma^+$ under the same field amplitude but opposite direction. Compared to the zero-field case, more optical transitions emerge above 1.8 eV and they blue shift with magnetic field. These features are identified as the exciton excited states or Rydberg states 2s, 3s, 4s, etc. The highest excited state that is identifiable at 9 T is 11s. This is an indication of excellent sample quality. For a given field, the energy difference for each exciton state probed by the $\sigma^+$ and $\sigma^-$ light corresponds to the exciton valley Zeeman splitting [29], which depends on field linearly. The Zeeman contribution can be eliminated by averaging of the two values [29].

At the Dirac point ($V_g \approx 0$ V), graphene has the minimum screening effect on the nearby WSe$_2$ layer, and the Rydberg states in WSe$_2$ retain substantial binding energies. The 1s and 2s excitons are well resolved and have approximately linear field dependence for the entire range of field from -9 T and 9 T. The exciton excited states 3s, 4s, … become better resolved at higher fields. Their field dependence evolves from being quadratic (from the exciton diamagnetic shift) to linear, depending on whether the exciton Bohr radius or the magnetic length is the dominant length scale [29]. The behavior observed here is qualitatively similar to previous studies on monolayer transition metal dichalcogenides (TMDs) in a uniform dielectric environment [29,39]. We employ similar methods to determine the quasiparticle band gap $E_g$ and exciton binding energies in the following.

For monolayer TMD semiconductors in a uniform dielectric environment of relative permittivity $\kappa$, the electron-hole interactions are well described by the Keldysh potential [40]:

$$V_K(r) = -\frac{e^2}{8\epsilon_0 r_0}[H_0\left(\frac{\kappa r}{r_0}\right) - Y_0(\frac{\kappa r}{r_0})], \qquad (1)$$

where $r$ is the electron-hole separation, $r_0$ is the characteristic screening length, and $H_0$ and $Y_0$ are the Struve and Bessel functions of the second kind, respectively. Under an out-of-plane magnetic field $B$, the low-energy Hamiltonian of the electron-hole pair can be written as:

$$H^{(s)} = -\frac{\hbar^2}{2m_r}\left(\partial_r^2 + \frac{1}{r}\partial_r\right) + \frac{e^2 B^2}{8m_r}r^2 + V_K(r) \tag{2}$$

for the axially symmetric s-states. Here $\hbar$ is the reduced Planck constant and $m_r$ is the reduced mass of the exciton. The Zeeman terms are neglected.

We numerically solve the Schrödinger equation with Hamiltonian given by Eqn. (2) and compare the energy of the exciton states to the average of the measured ones with the $\sigma^+$ and $\sigma^-$ probes. We use 1000 grids and a maximum distance of 200 nm in the numerical calculations. The following parameters are found to describe the experiment at $V_g \approx 0$ V the best: $m_r = 0.16\, m_0$ ($m_0$ denoting the free electron mass), $\kappa = 6.7$, $r_0 = 2.5$ nm, and $E_g = 1.833$ eV. The calculated exciton energies are plotted as dashed red curves in Fig. S4b. In comparison, $\kappa$ is reported to be 4.5 for a TMD monolayer encapsulated in hBN without graphene [29]. Our result shows that a single layer of graphene even at its Dirac point can significantly enhance the dielectric screening effect in WSe$_2$ when it is placed immediately above it.

Away from the graphene Dirac point (e.g. $V_g = 5$ V), the exciton excited states in WSe$_2$ can no longer be clearly identified at $B = 0$ T. Compared to the case of $V_g \approx 0$ V, more optical transitions develop under magnetic fields. These states are more compact and disperse approximately linearly in field. We have attempted to use the same Hamiltonian of Eqn. (2) to fit the data. However, no parameters can be identified to capture all exciton states. We plot the calculated magnetic-field dependence of the exciton energies with $m_r = 0.17 m_0$, $\kappa = 19.5$, $r_0 = 2.5$ nm, and $E_g = 1.8$ eV in Fig. S5b. The parameters are chosen to best describe the exciton excited states (2s, 3s, …, and 8s). The predicted 1s exciton energy is about 60 meV above the experimental value.

The failure of the Keldysh potential for our device geometry is not surprising since the screening effect from a single layer of graphene cannot be well described by a uniform dielectric environment with an effective $\kappa$. This becomes a particular issue when graphene is doped and the screening effect is strong. In addition, the finite sample-gate distance (about 30 nm) needs to be properly taken into account, for instance, by considering the image charge effect from the gate electrode. However, a full theoretical description of the device is beyond the scope of the current work and future studies are required.

**Determination of the quasiparticle band gap**
We determine the gate-dependent quasiparticle band gap or the band-edge transition energy $E_g$ in monolayer WSe$_2$ from the optical reflection contrast spectrum. We divide the analysis into two regimes, close to the graphene Dirac point or the secondary Dirac points and away from the Dirac points. Below we show examples with $V_g \approx 0$ V and 5 V from device D1 to represent the two regimes. We have used similar methods to determine the transition energy between moiré bands in devices that exhibit the optical transition replicas.

Near the Dirac point ($V_g \approx 0$ V), the screening effect is relatively weak and the Rydberg states in WSe$_2$ retain substantial binding energies. Up to three exciton states can be identified in the reflection contrast spectrum at zero magnetic field (Fig. 2b). We evaluate $E_g$ from the 2s and 3s spacing based on the 2D hydrogen model. In this model, the $N$s exciton binding energy is given by $E_b^{(Ns)} = \frac{m_r e^4}{2\hbar^2 (4\pi\kappa\varepsilon_0)^2 (N-\frac{1}{2})^2} \propto \frac{1}{(N-\frac{1}{2})^2}$ ($N$ = 1, 2, etc). The 2s exciton binding energy $E_b^{(2s)}$ is related to the 2s and 3s spacing $\delta^{(2s,3s)}$ by a constant factor, $E_b^{(2s)} = \frac{25}{16} * \delta^{(2s,3s)}$. The quasiparticle band gap energy can then be evaluated from experiment as $E_g = E^{(2s)} + E_b^{(2s)}$, where $E^{(2s)}$ is the 2s exciton energy. From experiment we obtain $\delta^{(2s,3s)} = 10$ meV, $E_b^{(2s)} = 15.6$ meV, and $E_g = 1.832$ eV. The gap value agrees well with that from the analysis of the Keldysh potential model described in S4. The Keldysh potential approaches the unscreened Coulomb potential for large electron-hole separations ($r > r_0$). The solution to the simple 2D hydrogen model becomes increasingly accurate for high-energy exciton states with increasingly large exciton Bohr radii.

Away from the Dirac points ($V_g = 5$ V), the screening effect increases and the exciton excited states are no longer identifiable at zero magnetic field. We extract the quasiparticle band gap energy from the magneto-optical measurement. Figure S5c shows the energy of the $N$s state as a function of $N$ for $N$ = 2 to 11 at $B$ = -9 T. It follows a linear dependence for $N > 4$ with spacing between the exciton states approaching the exciton cyclotron energy $\delta_B = \hbar e B / m_r \approx 6.3$ meV. The 2s and 3s spacing $\delta^{(2s,3s)}$ deviates from the linear dependence by less than 1 meV. We therefore estimate the 2s exciton binding energy $E_b^{(2s)}$ to be ~ 1 meV based on the 2D hydrogen model. The binding energy of the higher excited states is expected to diminish rapidly. When the exciton binding energies are much smaller than $\delta_B$, the exciton transitions evolve into the fan-like interband Landau level transitions. They depend on field approximately linearly and converge to a single energy 1.8 eV at zero field. We assign this energy the quasiparticle band gap.

Figure S5d shows the reflection contrast spectrum at zero field with the band edge transition from above analysis marked by a dashed vertical line. The band edge transition corresponds to the high-energy tail of a hump feature. In the figure we also show the derivative of the reflection contrast spectrum. The band edge transition can be identified as a crossover from negative values for the derivative to zero. This provides a quick estimate of $E_g$ without performing a systematic magneto-optic measurement. The uncertainty in the gap energy is estimated to be < ~1 meV. We show the result of $E_g$ estimated using this method (black curve) as a function of gate voltage (except the Dirac point) for device D1 in Fig. S6. The value at the Dirac point is from the 2D hydrogen model discussed above. The color background is the first derivative of the reflection contrast spectrum with respect to energy at zero magnetic field.

The band gap energy $E_g$ decreases rapidly with increasing gate voltage (also shown in Fig. 2c). It is more informative to show $E_g$ as a function of $|n|^{1/2}$ (inset of Fig. 2c). Away from the Dirac point, $E_g$ depends nearly linearly on $|n|^{1/2}$. Although renormalization of the quasiparticle gap by dynamical screening is a challenging theoretical problem, the observed doping dependence of $E_g$ can be qualitatively understood by invoking the doping-dependent complex dielectric function of graphene $\kappa(k,\omega)\varepsilon_0 = \varepsilon_b + \frac{2\pi e^2}{k}\Pi(k,\omega)$. It depends on both the wavevector $k$ and the frequency $\omega$. Here $\varepsilon_b$ is the doping-independent background dielectric constant of graphene and $\Pi(k,\omega)$ is the 2D electronic polarizability. In the long-wavelength limit ($k \to 0$), $\Pi(k,\omega)$ is proportional to the electronic density of states of graphene, which is proportional to $|n|^{1/2}$ (ref.[31]). The real part of $\Pi(k,\omega)$ is expected to give rise to a quasiparticle energy shift and the imaginary part to increased damping. Our system belongs to this regime when away from the graphene Dirac points since the graphene Fermi energy (which can be easily larger than 100 meV) is much larger than the binding energy of the exciton excited states. The graphene dielectric function is proportional to $|n|^{1/2}$. The quasiparticle energy shift is thus proportional to $|n|^{1/2}$ to the first order approximation.

**References for Methods**:

**Figures and figure captions:**

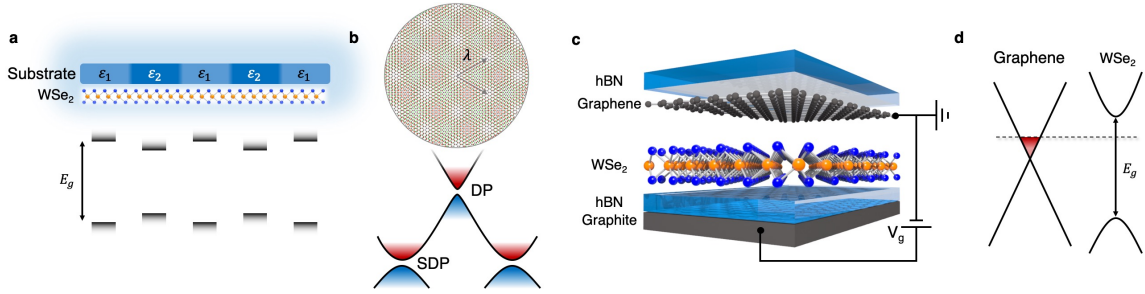

**Figure 1. Illustration of creating spatially periodic electronic band structure in monolayer WSe₂ by dielectric screening.** (**a**) A spatially modulated quasiparticle gap ($E_g$) is formed in monolayer WSe₂ on a substrate with periodic dielectric constant ($\varepsilon_2 > \varepsilon_1$). (**b**) Top: graphene/hBN moiré superlattice of periodicity $\lambda$; Bottom: the corresponding modified graphene band structure, where gap opens at the Dirac point (DP) and secondary Dirac points (SDPs). (**c**) Schematic of a typical device structure. From the bottom up, it consists of a graphite gate electrode, hBN gate dielectric, WSe₂ monolayer, and a graphene layer on hBN substrate. Some devices also contain a thin hBN spacer that separates the WSe₂ and graphene layer and is not shown. To tune the device doping density, a gate voltage $V_g$ is applied to the graphite gate electrode while the graphene layer is grounded. (**d**) Band alignment between graphene and monolayer WSe₂. The DP is deep inside the WSe₂ band gap. The applied gate voltage shifts the graphene Fermi level (dashed line) while WSe₂ remains charge neutral.

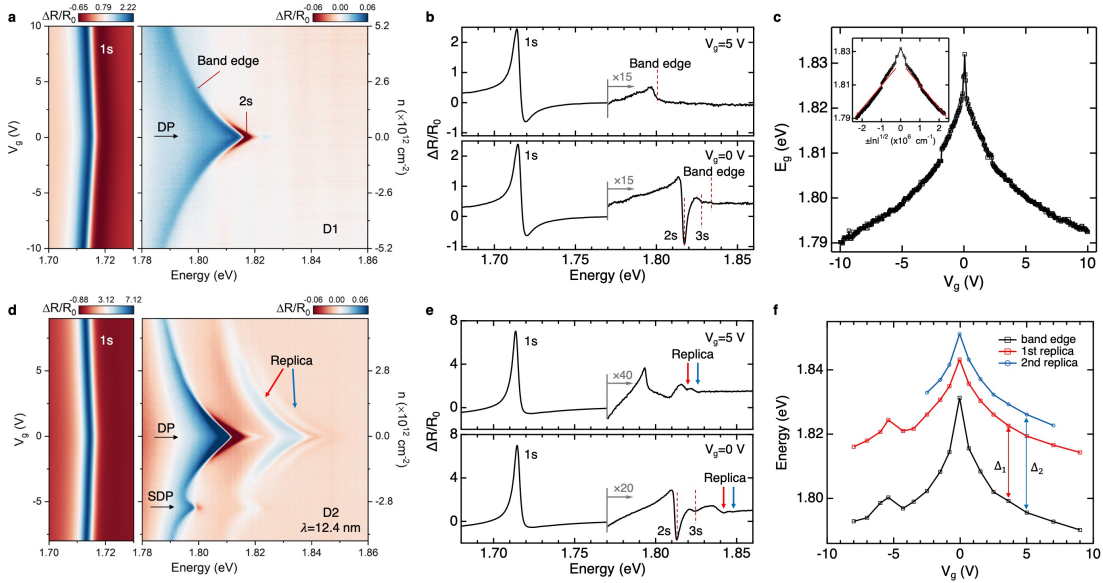

**Figure 2. Reflection contrast of devices without and with graphene/hBN alignment.** (**a**) Gate-dependent reflection contrast ($\Delta R/R_0$) spectrum of device D1 with misaligned graphene on hBN. The horizontal arrow denotes the gate voltage for the graphene DP. (**b**) Representative linecuts of (**a**) at $V_g = 5$ V and 0 V. The spectra above 1.77 eV are multiplied by a factor for clarity. The dashed red lines mark the 2s, 3s and band edge transitions. (**c**) The extracted quasiparticle band gap $E_g$ of monolayer WSe$_2$ in device D1 as a function of $V_g$ and $\pm|n|^{1/2}$ (inset), where $n$ is the doping density and the + (-) sign denotes electron (hole) doping. The red lines show a linear dependence of $E_g$ on $|n|^{1/2}$ away from the DP. (**d,e**) Same as **a**, **b** for device D2 with graphene/hBN superlattice period $\lambda = 12.4$ nm. The satellite feature in **d** is the graphene SDP. The red and blue arrows in **d**, **e** indicate the optical transition replicas. (**f**) Transition energy of the fundamental band edge ($E_g$) and two replicas of monolayer WSe$_2$ in device D2. Away from the DPs the replicas are at energy $\Delta_1$ and $\Delta_2$ above the fundamental band edge transitions.

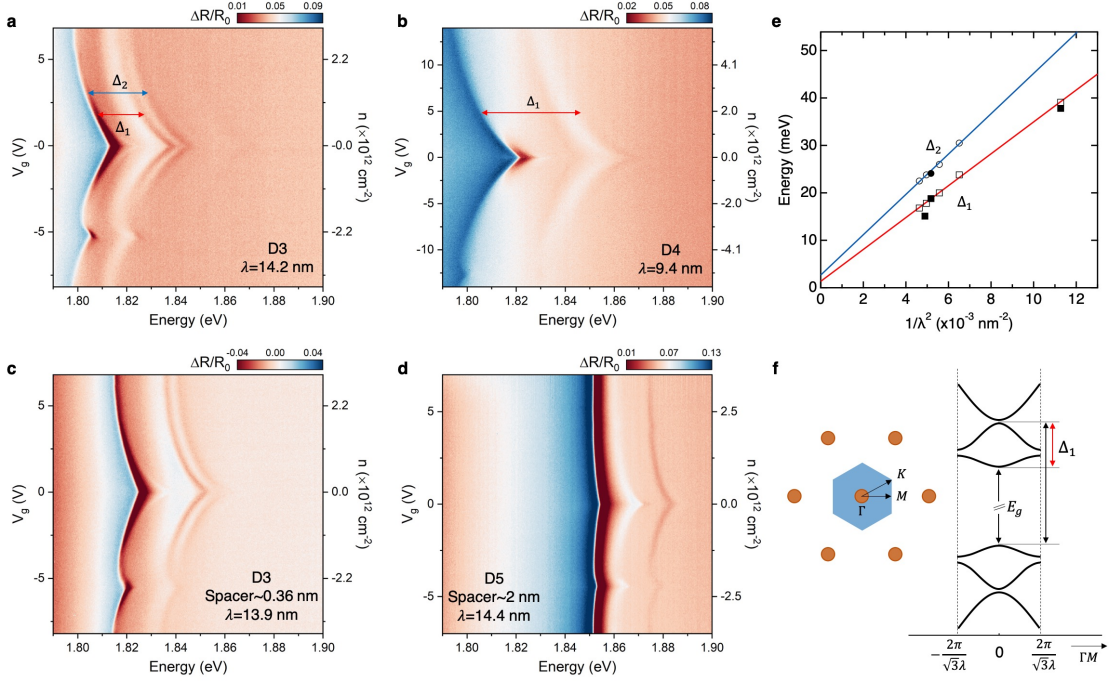

**Figure 3. Effects of superlattice period and spacer thickness**. (**a-d**) Gate-dependent reflection contrast spectrum for one half of device D3 (superlattice period $\lambda$ = 14.2 nm) with no spacer (**a**) and the other half with a ~ 0.36-nm hBN spacer (**c**), device D4 ($\lambda$ = 9.4 nm) with no spacer (**b**), and device D5 ($\lambda$ = 14.4 nm) with a 2-nm hBN spacer (**d**). Away from the Dirac points, the replicas are $\Delta_1$ and $\Delta_2$ (denoted by the red and blue arrows, respectively) above the fundamental band edge transitions. Only one optical transition replica is observed in some devices, reflecting a weaker moiré potential. (**e**) Dependence of $\Delta_1$ (squares) and $\Delta_2$ (circles) on $1/\lambda^2$ for all measured devices that exhibit the optical transition replicas. The open (filled) symbols represent devices without (with) a spacer. The solid lines are linear fits of $\Delta_2$ and $\Delta_1$ considering only the open symbols. The ratio of the two slopes is about 1.3. (**f**) Left: mini Brillouin zone formed in WSe$_2$ due to Bragg scattering from the spatially periodic dielectric screening effect. Right: band folding at the mini Brillouin zone boundaries along the $\Gamma M$ direction. The folded bands avoid crossing and allow new optical transitions at $\Delta_1$ from the original band edge (first replica). The second replica could arise from similar effects due to zone folding along the $\Gamma K$ direction.

# Supplementary Information

**Supplementary data and figures**

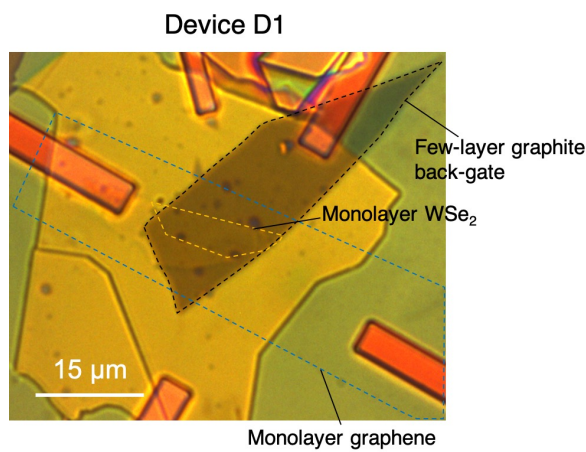

Fig. S1. **Optical image of device D1.** The black, blue, and yellow dashed curves outline the few-layer graphite gate electrode, the graphene layer, and the WSe$_2$ monolayer, respectively. The scale bar is 15 $\mu$m. The device structure of D1 is shown in Fig. 1**c**.

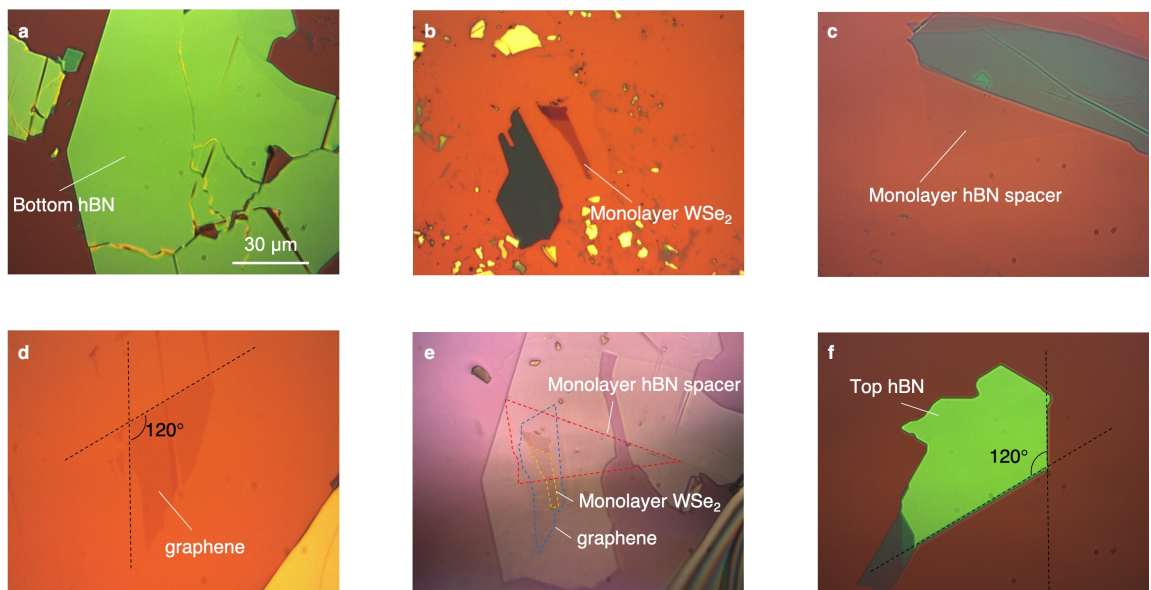

Fig. S2. **Optical images from device D3**. (**a-d**, **f**) Components of device D3 on Si substrates before transfer. They include, from the bottom up according to Fig. 1**c**, the bottom hBN layer (**a**), the WSe$_2$ monolayer (**b**), the monolayer hBN spacer (**c**), the graphene layer (**d**), and the top hBN layer (**f**). (**e**) An optical image of the sample before picking up the top hBN layer. The WSe$_2$ monolayer, the one-layer hBN spacer, and the graphene layer are outlined with yellow, red, and blue dashed curves, respectively. The sharp edge (120° angle) of the graphene layer (**d**) and the top hBN layer (**f**) are aligned

before transfer. The final device shows optical transition replicas in areas both with (Fig. 3**c**) and without the hBN spacer (Fig. 3**a**).

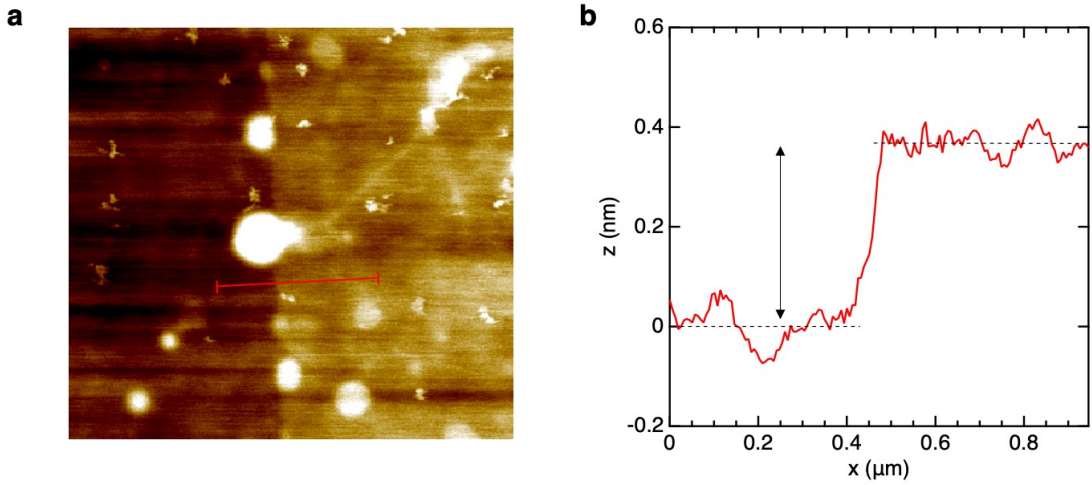

Fig. S3. **AFM of device D3 (part of the sample).** (**a**) AFM topography image. (**b**) Height measurement along the red line in (**a**). The step size (~ 0.36 nm) corresponds to the thickness of an hBN monolayer.

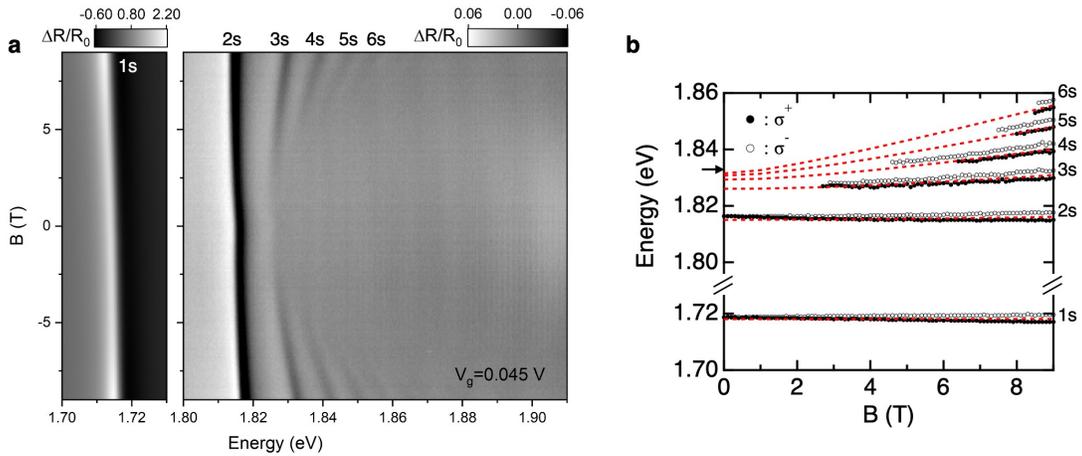

Fig. S4. **Magneto-optical spectroscopy of device D1 at $V_g \approx 0$ V**. (**a**) Reflection contrast spectrum ($\Delta R/R_0$) as a function of out-of-plane magnetic field at the graphene Dirac point. The white light probe is left circularly polarized ($\sigma^+$). (**b**) The magnetic-field dependence of the 1s, 2s, 3s, …6s exciton energy is extracted from (**a**). The filled and empty symbols denote the values measured with $\sigma^+$ and $\sigma^-$ light, respectively. The latter is equivalent to $\sigma^+$ probe under a negative magnetic field. The red dashed lines are best fits of the Keldysh potential model as described in the text.

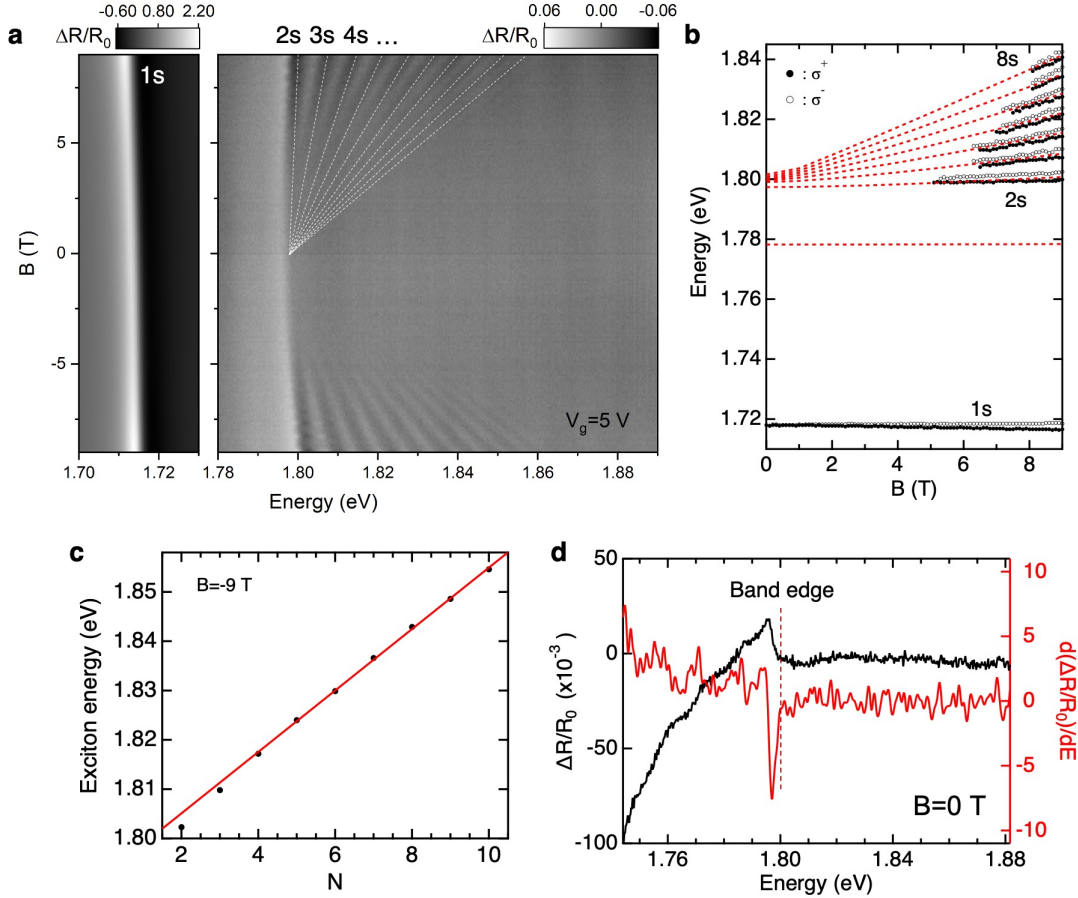

Fig. S5. **Magneto-optical spectroscopy of device D1 at $V_g$ = 5 V**. (**a**, **b**) Same as Fig. S4 for doped graphene. The dashed lines in (**a**) show the fan-like interband Landau level transitions for $B > 0$ T, which converge to the band gap energy at zero field. Unlike the case at the graphene Dirac point, the Keldysh potential model cannot describe all the exciton states. The fitting parameters are chosen to best match the exciton excited states. The predicted 1s energy from the model is about 60 meV higher than the experimental result. (**c**) Energy of the $N$s state as a function of $N$ for $N = 2 – 11$ at $B = -9$ T. Solid red line is a linear fit. (**d**) Reflection contrast spectrum (left, black line) and its first derivative with respect to energy (right, red line) at zero magnetic field. The vertical red dashed line indicates the band edge transition (~ 1.8 eV).

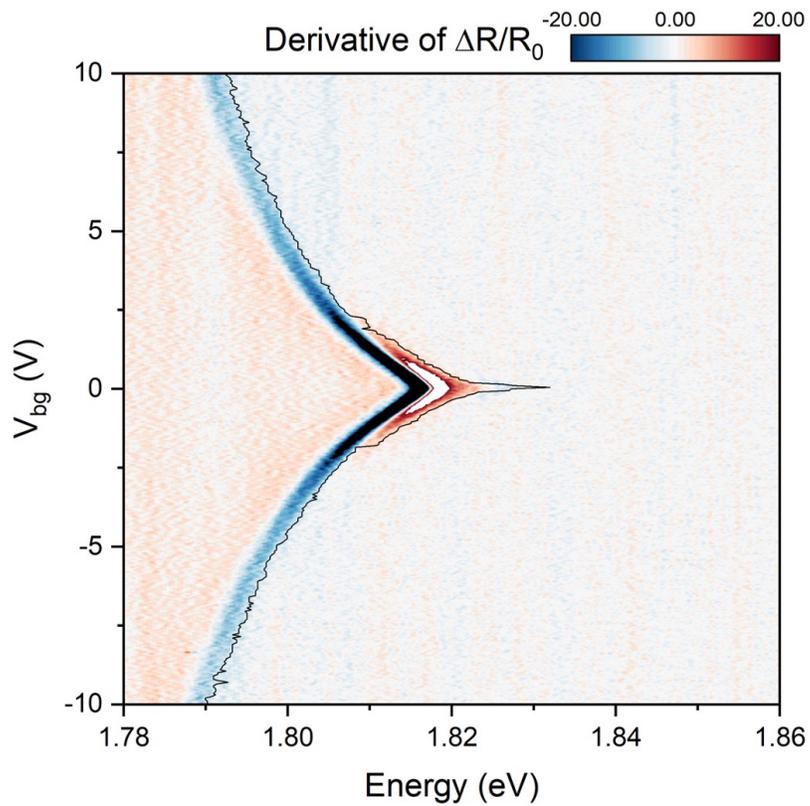

Fig. S6. **Gate-dependent quasiparticle band gap energy of device D1 at zero magnetic field.** The contour plot is the first derivative of the reflection contrast spectrum with respect to energy. The black curve is the gate-dependent band edge transition determined using the methods described in the text.